\documentclass[preprint,12pt]{aastex}

\usepackage{epstopdf}

\usepackage{graphicx}

\usepackage{subfigure}

\begin{document}




\def\putplot#1#2#3#4#5#6#7{\begin{centering} \leavevmode

\vbox to#2{\rule{0pt}{#2}}

\includegraphics{#1}





\end{centering}}




\def\Msun{M_\odot}

\def\Lsun{L_\odot}

\def\Rsun{R_\odot}

\slugcomment{Submitted to ApJ}

\shorttitle{T Pyx Shell}

\shortauthors{Shara et al}

\title{{\it HST} Images Flash Ionization of Old Ejecta by the 2011 Eruption of Recurrent Nova T Pyxidis\altaffilmark{1}}

\author{Michael~M.~Shara\altaffilmark{2}, David~Zurek\altaffilmark{2}, Bradley~E.~Schaefer\altaffilmark{3}, Howard E. Bond\altaffilmark{4,5},  Patrick Godon\altaffilmark{6}, Mordecai-Mark~Mac Low\altaffilmark{2}, Ashley Pagnotta\altaffilmark{2}, Dina~Prialnik\altaffilmark{7}, Edward M. Sion\altaffilmark{6}, Jayashree Toraskar\altaffilmark{2}, and Robert E. Williams\altaffilmark{5}}





\altaffiltext{1}{Based on observations with the NASA/ESA {\it Hubble Space
  Telescope}, obtained at the Space Telescope Science Institute, which is
  operated by AURA, Inc., under NASA contract NAS 5-26555.}

\altaffiltext{2}{Department of Astrophysics, American Museum of Natural History, Central Park West and 79th Street, New York, NY 10024-5192}

\altaffiltext{3} {Department of Physics and Astronomy, Louisiana State University, 243 Nicholson, Baton Rouge, LA 70803-0001}

\altaffiltext{4} {Department of Astronomy \& Astrophysics, Pennsylvania State University, University Park PA 16802}

\altaffiltext{5} {Space Telescope Science Institute, 3700 San Martin Drive, Baltimore MD 21218}

\altaffiltext{6} {Department of Astrophysics and Planetary Science, Villanova University, Villanova, PA 19085}

\altaffiltext{7} {Department of Geophysics and Planetary Sciences, Tel-Aviv University, Ramat Aviv, Israel}

\begin{abstract}

T Pyxidis is the only recurrent nova known to be surrounded by knots of material ejected in previous outbursts. Following the eruption that began on 2011 April 14.29, we obtained seven epochs (from 4 to 383 days after eruption) of {\it Hubble Space Telescope} narrowband H$\alpha$ images of T Pyx . The ionizing flash of radiation from the nova event had no discernible effect on the surrounding ejecta until at least 55 days after the eruption began. Photoionization of hydrogen located north and south of the central star was seen 132 days after the beginning of the eruption. That photoionized hydrogen recombined in the following 51 days, allowing us to determine a hydrogen atom density of at least $7\times 10^{5} \, \rm cm^{-3}$ - at least an order of magnitude denser than the previously detected, unresolved [\ion{N}{2}] knots surrounding T Pyx. Material to the northwest and southeast was photoionized, and became bright between 132 and 183 days after the eruption began. 99 days later that northwest and southeast hydrogen had recombined. Both then (282 days after outburst) and 101 days later, we detected almost no trace of hydrogen emission around T Pyx. We determine that there {\it is} a large reservoir of previously unseen, cold diffuse hydrogen overlapping the previously detected, [\ion{N}{2}] - emitting knots of T Pyx ejecta. The mass of this newly detected hydrogen is model-dependent, but is is probably an order of magnitude larger than that of the [\ion{N}{2}] knots. We also determine that there is {\it no} significant reservoir of undetected hydrogen-rich ejecta, with density comparable to the flash-ionized ejecta we have detected, from the outer boundaries of the previously detected ejecta out to about twice that distance. The lack of distant ejecta is consistent with the Schaefer et al (2010) scenario for T Pyx, in which the star underwent its first eruption within five years of 1866 after many millennia of quiescence, followed by the six observed recurrent nova eruptions since 1890. The lack of distant ejecta, demonstrated by our observations, is {\it not} consistent with scenarios in which T Pyx has been erupting continuously as a recurrent nova for many centuries or millennia. 

\end{abstract}


\section{Introduction and Motivation}

Classical novae are cataclysmic binary stars, wherein a white dwarf (WD) accretes a hydrogen-rich envelope from its Roche-lobe filling companion, or from the wind of a nearby giant. Theory \citep{sha81} and detailed simulations \citep{yar05} predict that once the pressure at the base of the accreted envelope of the WD exceeds a critical value, a thermonuclear runaway (TNR) will occur in the degenerate layer of accreted hydrogen. The TNR causes the rapid rise of the WD's luminosity to $\sim 10^{5} L_\sun$ or more, and the high-speed ejection of the accreted envelope \citep{sst72, pss78, sha89} in a classical nova explosion. 

Recurrent novae (RNe) are a subclass of cataclysmic binaries, brightening dramatically for a few weeks at intervals of decades \citep{web87}.  Comprehensive photometric histories of all ten known Galactic RNe are detailed in \citet{sch10}. Although RN outbursts are powered by TNRs resembling those of classical novae, the latter have eruptions separated by thousands of years instead of decades \citep{pat84,sha86,sha12}.

The RN T Pyxidis, surrounded by well-resolved ejecta, is an important test case for studying the eruptions and ejecta of all classical novae. Before 2011, T Pyx had five recorded outbursts to its credit (1890, 1902, 1920, 1944, 1966), all of which were very similar photometrically \citep{may67}. Its eruptions every few decades have driven successive blast waves out into the ejecta around the RN and, presumably, the interstellar medium beyond. Successive ejection events must result in collisions and interactions between their ejecta. Simulations show that these collisions lead to Rayleigh-Taylor instabilities that fragment the ejecta, producing the observed, elongated structures \citep{tor13}. \citet{con97} demonstrated that when two ejected nova shells collide, a primary blast wave propagates outward into the older ejecta, while a reflected reverse shock wave propagates back through the new ejecta. Different spectral lines are emitted in the two zones, and combined, the \citet{con97} model accounts well for the observed line spectrum of the T Pyx ejecta. 

Narrowband ground-based H$\alpha$ imaging of T~Pyx by \citet{due79} and \citet{wil82} revealed a shell with diameter $\sim$$10''$ surrounding the star. \citet{smw89} obtained deeper ground-based CCD images of T~Pyx that revealed a faint extended H$\alpha$ + [N~II] halo twice as large as the previously detected shell. Subsequently,  \citet{sha97} obtained {\it Hubble Space Telescope} ({\it HST}) Wide Field Planetary Camera 2 (WFPC2) images of T~Pyx in narrowband H$\alpha$ and [N~II] $\lambda$6584 filters, demonstrating  that (1)~the nebula is resolved into more than two thousand individual knots and (2)~the apparent expansion of knots away from the star is much slower than predicted from the radial velocities of ejecta \citep{cat69}.  

With a 13-year baseline, \citet{sps10} were able to detect the slow expansion of T Pyx's ejecta. The fractional expansion of the knots is constant, which implies no significant deceleration of the knots. They deduced that T~Pyx must have undergone an (unobserved) classical-nova eruption within $\pm5$ years of 1866, followed by six recurrent-nova outbursts. As T Pyx is the prototype of both short orbital-period and short recurrence-time Galactic RNe, it is important to independently check the \citet{sps10} evolutionary scenario. This is a key goal of the present paper. 

A rare opportunity for RN studies - a direct test of key predictions of the \citet{sps10} scenario and the \citet{tor13} hydrodynamics model - arose with T Pyx's most recent eruption, which began at 2011 April 14.29 = JD 2455665.79 \citep{waa11}. Exquisite light curves and spectroscopy of the eruption have been presented by \citet{sur14}.  \citet{che11}, \citet{sok13}, \citet{sho13}, and \citet{deg14} have investigated the three-dimensional geometry of the ejecta. \citet{che11} and \citet{sho13} have presented evidence that the ejecta are best represented by a bipolar model, oriented nearly face-on. The goal of \citet{sok13} was to image light scattered by dust; while the resulting images were of low S/N, they demonstrated that T Pyx's dust is concentrated in a clumpy ring about $5"$ in radius - - tilted 30 - 40$^\circ$ with respect to the plane of the sky - - with the eastern edge tilted toward the observer. The delay times between the direct optical light from the central source and the scattering of this light from dust in several clumps with the same foreground distance as the central source yielded a distance to T Pyx of  $4.8 \pm 0.5$  kpc. A completely independent distance estimate, based on the high radial velocity component of the Mg II doublet and the Galactic rotation curve in the direction of T Pyx, yielded an estimate of 5 kpc \citep{deg14}.

\citet{pat13} measured the period change across T Pyx's 2011 eruption, using it to deduce that more than $3 \times 10^{-5}M_\odot$ was ejected in that event. A similar conclusion was reached by \citet{nel14} on the basis of the high peak flux densities of T Pyx's radio emission, and by \citet{cho14} on the basis of the late appearance of supersoft X-rays. This much ejected mass is usually associated with a classical, rather than a recurrent nova. 

T Pyx is the only RN with clearly visible ejecta, and hence the only nova likely to illuminate its circumstellar ejecta, during the lifetime of {\it HST}. We therefore requested Director's discretionary time on {\it HST} with its Wide Field Camera 3 (WFC3) to image the T Pyx ejecta through the F656N narrowband H$\alpha$ filter. Our expectation was that, as the wave of intense nova illumination swept out into the nebula, hydrogen surrounding the nova would be photoionized and then recombine, brightening dramatically. The key questions we hoped to address with our observations were:

(1)~Is there (unexpected) matter outside the boundaries of the ejecta of the claimed 1866 classical nova eruption? The presence of significant amounts of such matter, illuminated by the nova flash, would cast significant doubt on the claim that an eruption within 5 years of 1866 was the first in many millennia: the basis of the evolutionary model of T Pyx \citep{sps10}. In particular, if T Pyx has been erupting for thousands of years, instead of just since 1866, there must be ejected material extending far beyond the boundaries of the knots imaged in \citet{sha97}.

(2)~Can we derive densities, and measure masses of T Pyx ejecta {\it not in} the prominent knots discovered by \citet{sha97}, from the lengths of time required for any observed flash-ionized material to fade? The extremely large measured masses of the 2011 eruption suggest that our previous determinations of ejecta mass, based on low density material emitting forbidden lines of [\ion{N}{2}], may be a significant underestimate. 

In Section 2 we describe our observations. We show narrowband H$\alpha$ imagery of the material surrounding T Pyx in Section 3. We use these images to test the \citet{sps10} scenario, and to demonstrate the presence of much previously invisible ejecta, surrounding T Pyx, in section 4. We briefly summarize our results in Section 5.   

\section {Observations}

\subsection{Ionization and Light-Echo Geometry} 

At a given time after the outburst, the illuminated material in the nebula surrounding T Pyx lies on a paraboloidal surface, the surface of constant light travel-time delay. This is strictly true for light scattered off dust.  An emission-line nebular knot, once ionized, will continue to radiate until it has recombined. Thus we only know that it is located between the parabola for which it was first detected and the next closer parabola at which it was not detected. This means that, for the highest-resolution 3D localization, the observations should be closely spaced in time. 

The unique relation between time delay and the 3D position of an illuminated knot in the nebula is illustrated in Figure 1,
which assumes a distance to T Pyx of 4.8 kpc \citep{sok13,deg14}. This figure 
shows how the paraboloids open up very quickly immediately after the outburst. It was thus essential to start the observations quickly in order to sample as much of the volume in front of the star as possible. Later observations were more widely spaced in time.  

\begin{figure}[ht]

\putplot{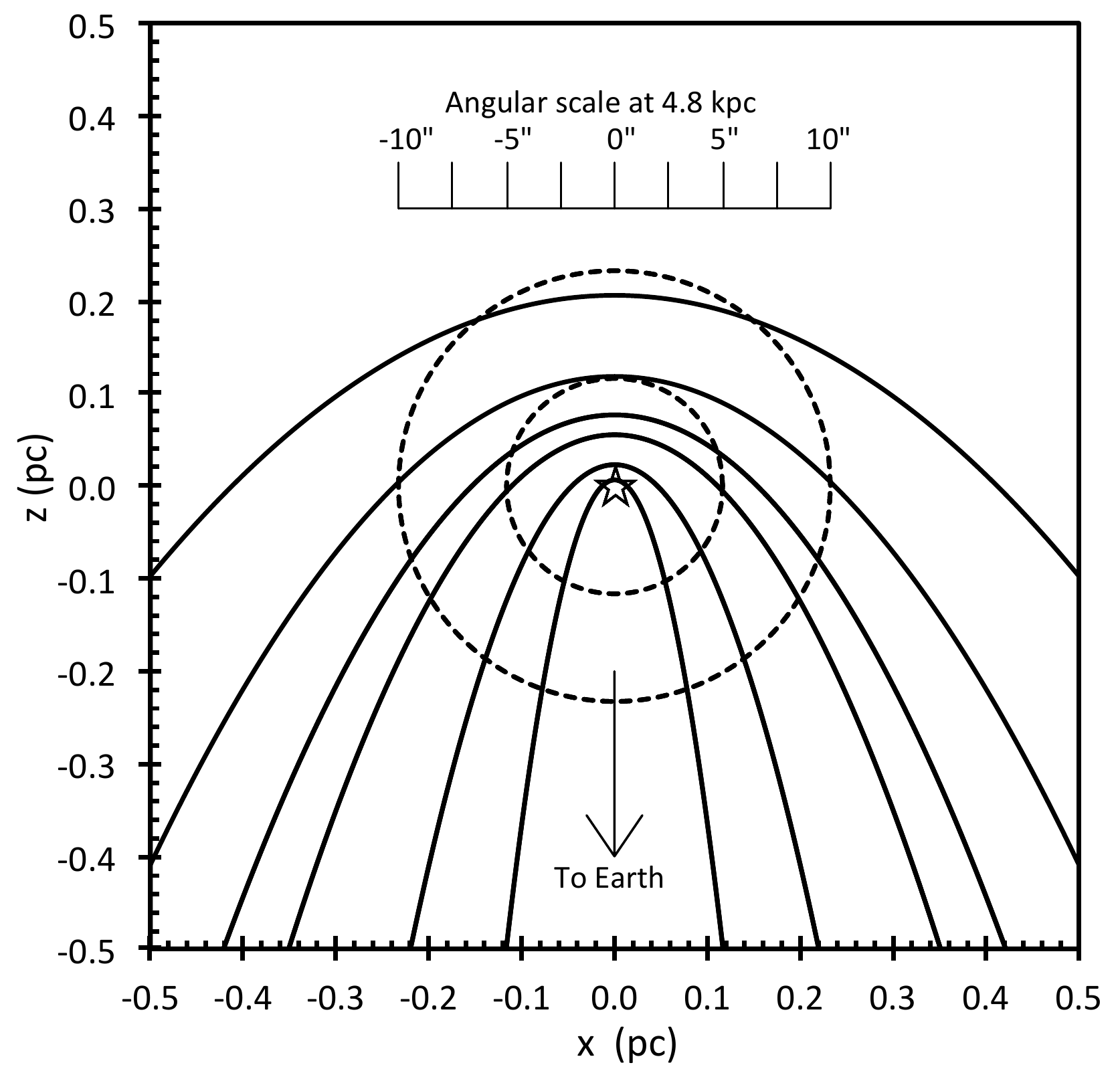}{.05 in}{0}{100}{100}{-200}{20}

\includegraphics[width=6in]{fig1.pdf}

\caption{Geometry of the ionization and light echoes expected from the 2011 April 14.29 outburst of T~Pyx. At a given time, the illuminated material lies on a paraboloid, with the open end pointing toward the Earth. In this figure, the parabolas are shown for six observation dates which are 16, 55, 132, 183, 282 and 383 days, respectively, 
after the beginning of the eruption on 2011 April 14.29. (The paraboloid corresponding to our first, filter calibration observation, just 4 days after the beginning of the eruption, is essentially a vertical line, and is not shown in the figure). The {\it angular\/} scale depends on the adopted distance to the star, here taken to be 4.8~kpc. 
The two dashed circles show the approximate angular sizes of the T~Pyx nebula, whose brightest portion has a radius of $\sim$$5''$, and whose
fainter portions extend out to $\sim$$10''$. Thus the illumination was expected to sweep through most of the brighter portions by about day 282, and should have
swept through the entire nebula shortly after day 383 (modulo the actual distance to T~Pyx; longer times for larger distances.)} 

\end{figure}

\clearpage

\subsection{The data}

Following its eruption on 2011 April 14.29, seven separate epochs of H$\alpha$ narrowband imagery of the recurrent nova T Pyx were obtained with the {\it HST} WFC3 \citep{kim08} for observing program GO-12446 (PI M. Shara). The camera provides a plate scale of $0.04"$ per pixel and a total field-of-view of $164"$ x $164"$. We used a 1K x 1K subarray of the camera to reduce readout time. Twelve H$\alpha$ images, each of 160 seconds duration, provided an effective exposure time of 1920 seconds at each epoch. Images were dithered during each epoch to allow for cosmic ray removal. 

The requested timings of the {\it HST} images were chosen based on the geometry of the expected ionization echoes and observational constraints on {\it HST} (cf Figure 1). We were aware that T Pyx near maximum light would saturate the CCD detector close to the central source, but nonetheless  requested that the first H$\alpha$ observation be made 4 days (2011 April 18) after the outburst began. This filter-calibration image (Figure 2) was intended to help us define the imaging characteristics (especially the scattered light or ``ghosts") within the F656N {\it HST} narrowband filter. Subsequently, images were obtained on 2011 April 30, 2011 July 08, 2011 September 05, 2011 November 15, 2012 February 22 and 2012 May 01, corresponding to 16, 55, 132, 183, 282 and 383 days, respectively, after the beginning of the eruption on 2011 April 14. Each of these six dates corresponds to an increasingly larger paraboloid in Figure 1. T Pyx was at visual magnitudes of about 7.5, 7.0, 8.8, 11.0, 11.8, 13.3 and 13.7 at the seven observation epochs noted above. The full light curve of T Pyx's 2011 eruption, with over 100,000 data points, is available on the American Association of Variable Star Observers website \footnote{http://www.aavso.org}.

\clearpage

\subsection{Image Processing}

We used the MultiDrizzle software package within PyRAF\footnote{PyRAF is a product of the Space Telescope Science Institute, which is operated by AURA for NASA.} to combine the dithered images from each epoch. Our science demanded the best-possible detections of the low surface brightness nebulosities surrounding T Pyx. Thus it was critical that the image subtractions we performed matched and removed the bright central pixels centered on T Pyx, as well as imaging artifacts (reflections, ghosts and diffraction spikes). 

Each of the seven epochs' images was taken with {\it HST} at a different roll angle, a natural consequence of maintaining {\it HST}'s solar panels normal to the Sun. This turned out to be beneficial because the diffraction spikes' and artifacts' locations remain fixed with respect to the CCD but rotate from epoch to epoch with respect to the sky. If each epoch's image is rotated to match the alignment of all the other epochs' images on the sky, the artifacts and diffraction spikes no longer align with respect to each other (while stars or nebulosity do align). It is essential to eliminate the diffraction spikes to detect low surface brightness nebulosities near T Pyx. 

We used epoch 1 (Figure 2) as a ``master calibration frame"  because no nebulosity is yet visible, while there was a strong signal in the diffraction spikes and other artifacts. As a first approximation we scaled stars near T Pyx in epoch 1 to the fluxes of the same stars in each later epoch for image differencing. 

The first five epochs were then processed as follows. We manually shifted epoch 1 (2011 April 18) in a grid of 2 by 2 pixels, in 0.1 pixel steps, with respect to the next four epochs' images. We scaled the image of epoch 1 to match the flux of each of the next four epochs, and subtracted each of those epochs from epoch 1. The resulting difference image with the smallest residual artifacts was then used as the center of a refined 0.3 by 0.3 pixel grid with 0.01 pixel steps to find the match that again minimized artifacts. 

The final shift we used was the one which left the smallest residuals for the diffraction spikes in each of the the difference images. This subtracted out the other artifacts very well, but not perfectly. The small residuals that can't be removed are due to small optical pathlength variations, a result of the telescope's temperature variations when entering and exiting the shadow of the earth, and jiitter, due to vibrations in {\it HST's} solar panels.

The flux in epochs 6 and 7 was so much lower than in the previous five epochs that the above procedure only added noise to these last two images. We therefore compared epochs 6 and 7 with each other. We repeated the process above but, instead of epoch 1, we used epoch 7 and shifted it in the drizzling process and scaled it to epoch 6. 
We checked that no constant features in epochs 6 and 7 cancelled each other out by performing the same subtraction after shifting the images with respect to each other (similar to ``unsharp masking" \citep{mal77} except that neither image was blurred). Except for two knots located 4 arc sec west of T Pyx, and noted below, no such features were detected.

\clearpage

\begin{figure}

\figurenum{2}

\epsscale{1.0}

\plotone{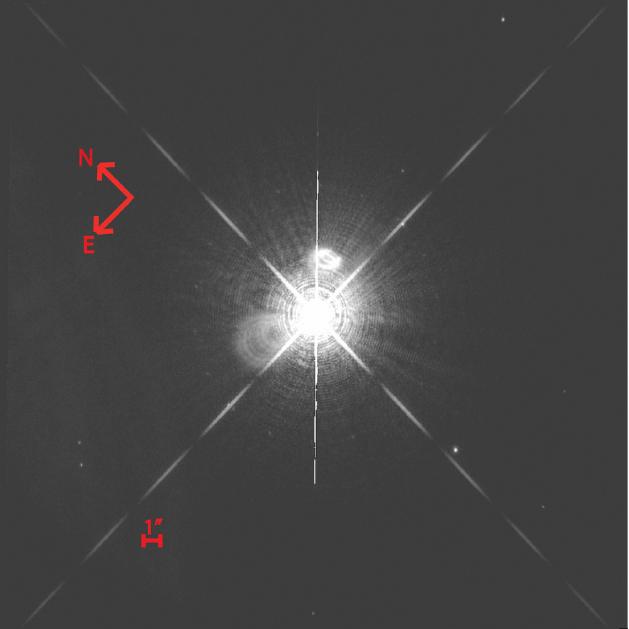}

\caption{A logarithmic stretch of the first H$\alpha$ image made with {\it HST's} WFC3 camera and F656N filter, 4 days (2011 April 18) after the latest outburst began. This image defines the imaging characteristics (especially the two strong, loop-shaped ``ghosts" of scattered light seen NE and NW of the saturated central star) of the F656N {\it HST} narrowband filter.}

\end{figure}

\clearpage

\section{The Images of T Pyx}

The dithered H$\alpha$ images of T Pyx at each of the six epochs after 2011 April 18 are shown as a mosaic in Figure 3. Not surprisingly, the glare of T Pyx and the ghosting in the H$\alpha$ filter dominate the images of 2011 April 30 and 2011 July 08; no unambiguous indication of nebulosity is seen. Eleven weeks later, on 2011 September 25, there {\it is} nebular matter clearly detected both north and south of T Pyx. This emission was extinguished, and was replaced by emission nebulosity to the northwest and southwest of the star just 51 days later, on 2011 November 15. Ninety-nine days later (2012 February 22) almost all traces of nebulosity were gone, and no new nebulosity appeared either then, or 101 days later (on 2012 May 01).

\clearpage

\begin{figure}

\figurenum{3}

\epsscale{1.0}

\plotone{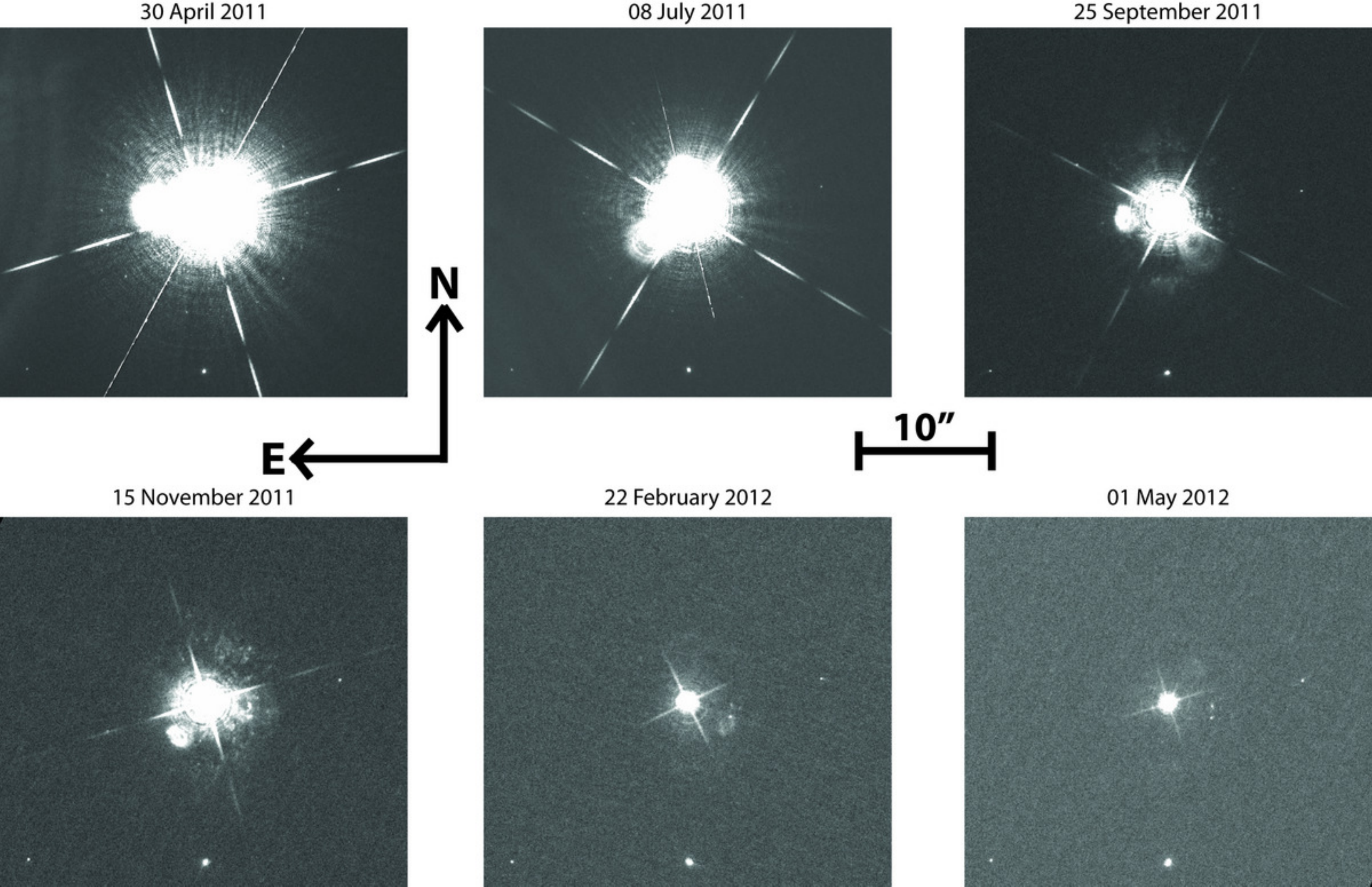}

\caption{Following Figure 2, these six {\it HST} F656N H$\alpha$ images of T Pyx were taken over an interval of 1 year. The images of the central source on 2011 April 30 and 2011 July 08 are saturated. Flash ionized hydrogen is present on 2011 September 25 and 2011 November 15, and mostly gone on 2011 February 22 and 2012 May 01. PSF-subtracted images (the following figure) make these changes more apparent.}

\end{figure}

\clearpage

The PSF-subtracted images used to remove much of the glare of T Pyx, and the F656N filter ghosting and diffraction spikes, are shown in Figures 4, 5, and 6. The nebulosities of 2011 September 25 (north and south of T Pyx) and 2011 November 15 (northwest and southwest of T Pyx) are much more apparent, as is the nearly complete lack of nebular emission on 2012 February 22 and 2012 May 10. 

To locate subtle edges and condensations we created unsharp mask images from our data by shifting and subtracting images from themselves.The F656N unsharp mask images (Figures 7 and 8) emphasize that much (though not all) of the newly detected, fluorescing hydrogen is diffuse, rather than being concentrated in knots. The comparison of the most recent epoch of [\ion{N}{2}] F658N images (from 2007) with the 2011 H$\alpha$ F656N images, shown in Figures 9 and 10, further emphasizes this difference in spatial distribution of the T Pyx ejecta.
\clearpage

\begin{figure}

\figurenum{4}

\epsscale{1.0}

\plotone{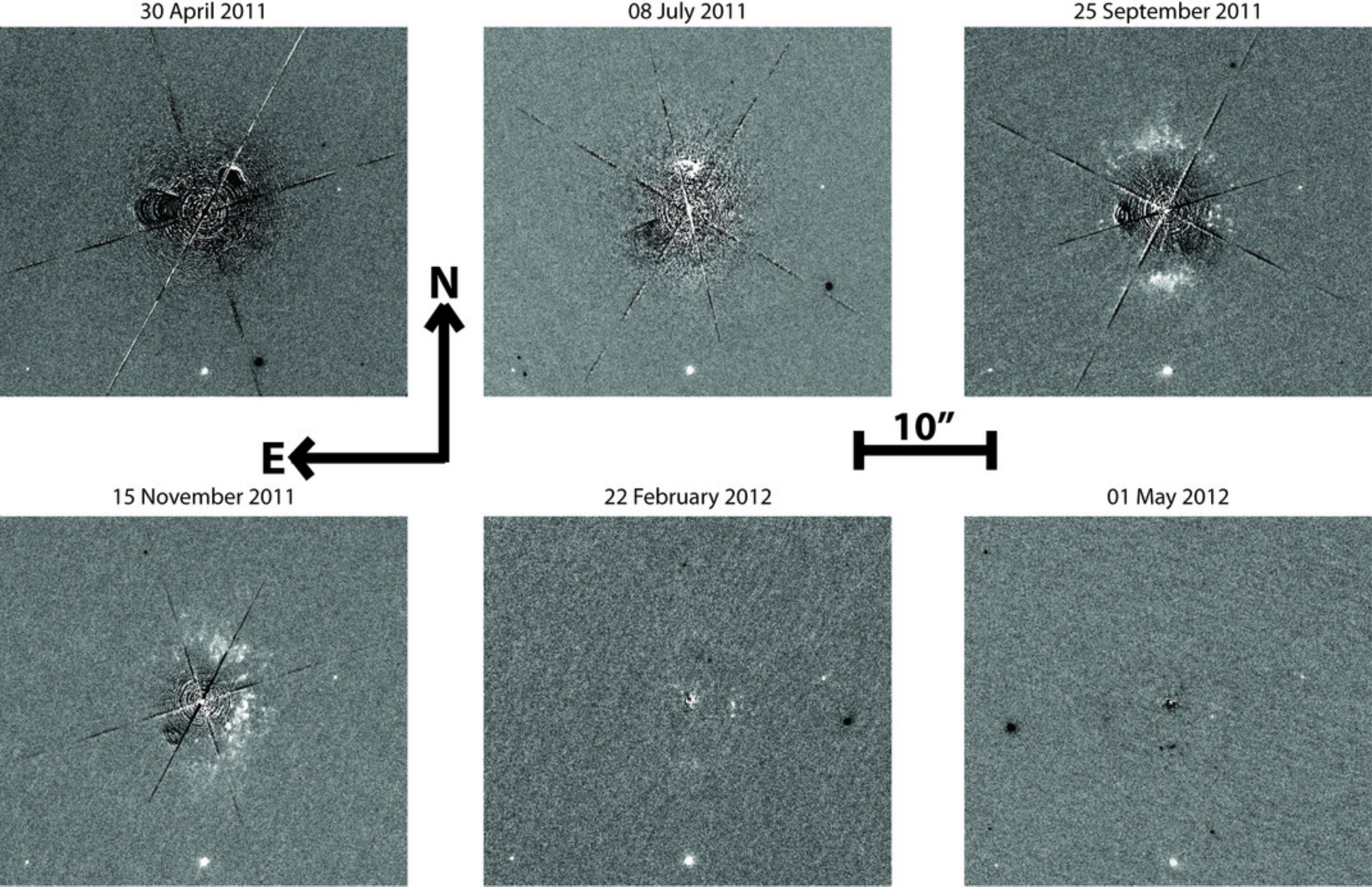}

\caption{F656N PSF-subtracted images of T Pyx taken by {\it HST} over the course of 1 year. The images in the first four epochs were subtracted from an appropriately scaled image shown in Figure 1. The last two images were subtracted from each other. See text for more details. The lobes seen on 2011 April 30 and 2011 July  08 are filter ghosts. Strong, flash-ionized H$\alpha$ emission is seen north and south of T Pyx on 2011 September 25 and northwest and southwest of the star on 2011 November 25. Only two faint, unresolved knots (4 arc sec west of T Pyx) remain on 2011 February 22 and 2012 May 01.}

\end{figure}

\clearpage

\begin{figure}

\figurenum{5}

\epsscale{1.0}

\plotone{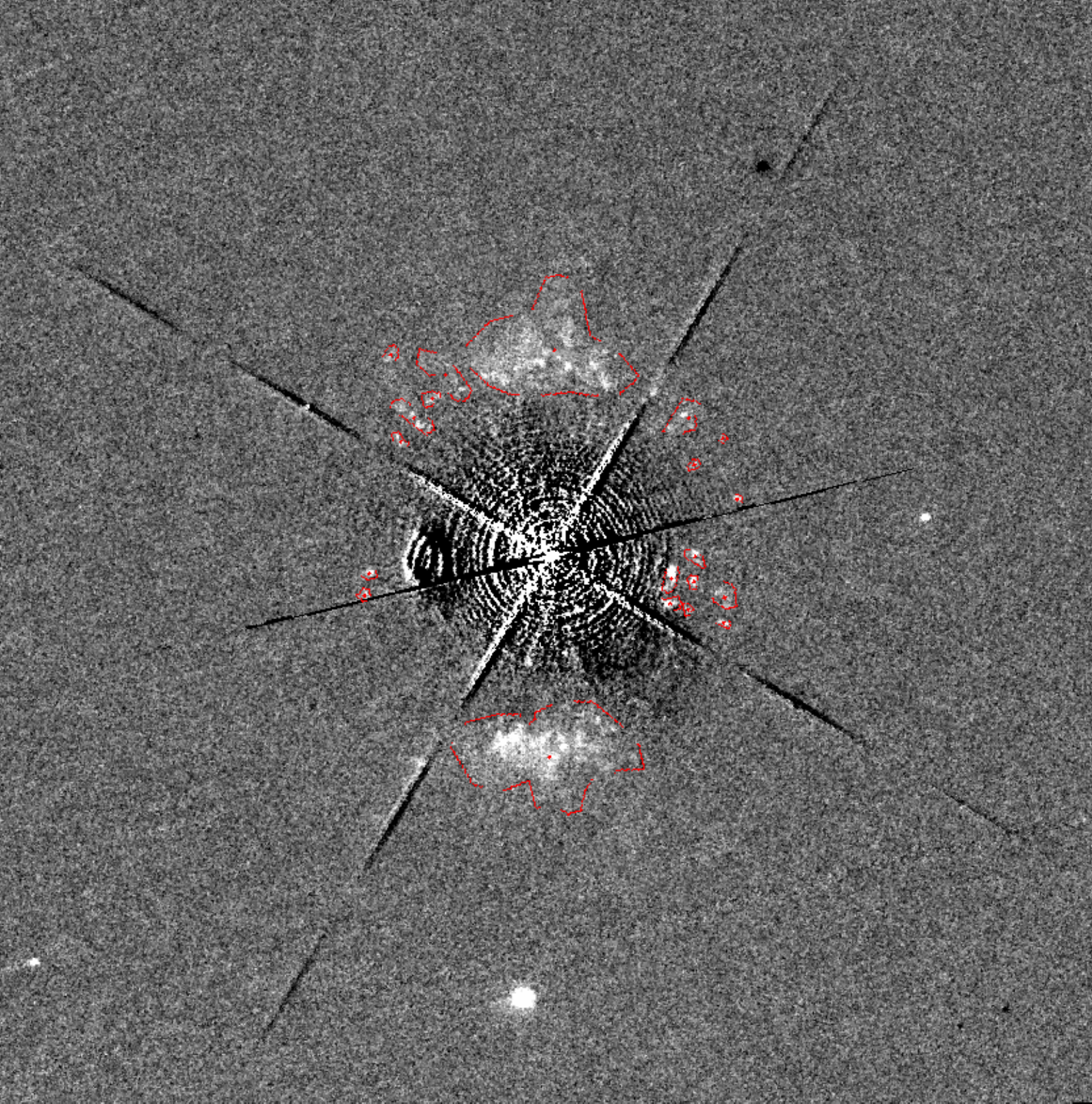}

\caption{Closeup view of the PSF-subtracted F656N images of T Pyx taken by {\it HST} on 2011 September 25. North is up, East is left and the image is $30"$ on each side. The encircled areas highlight the H$\alpha$ emission. }

\end{figure}

\clearpage

\begin{figure}

\figurenum{6}

\epsscale{1.0}

\plotone{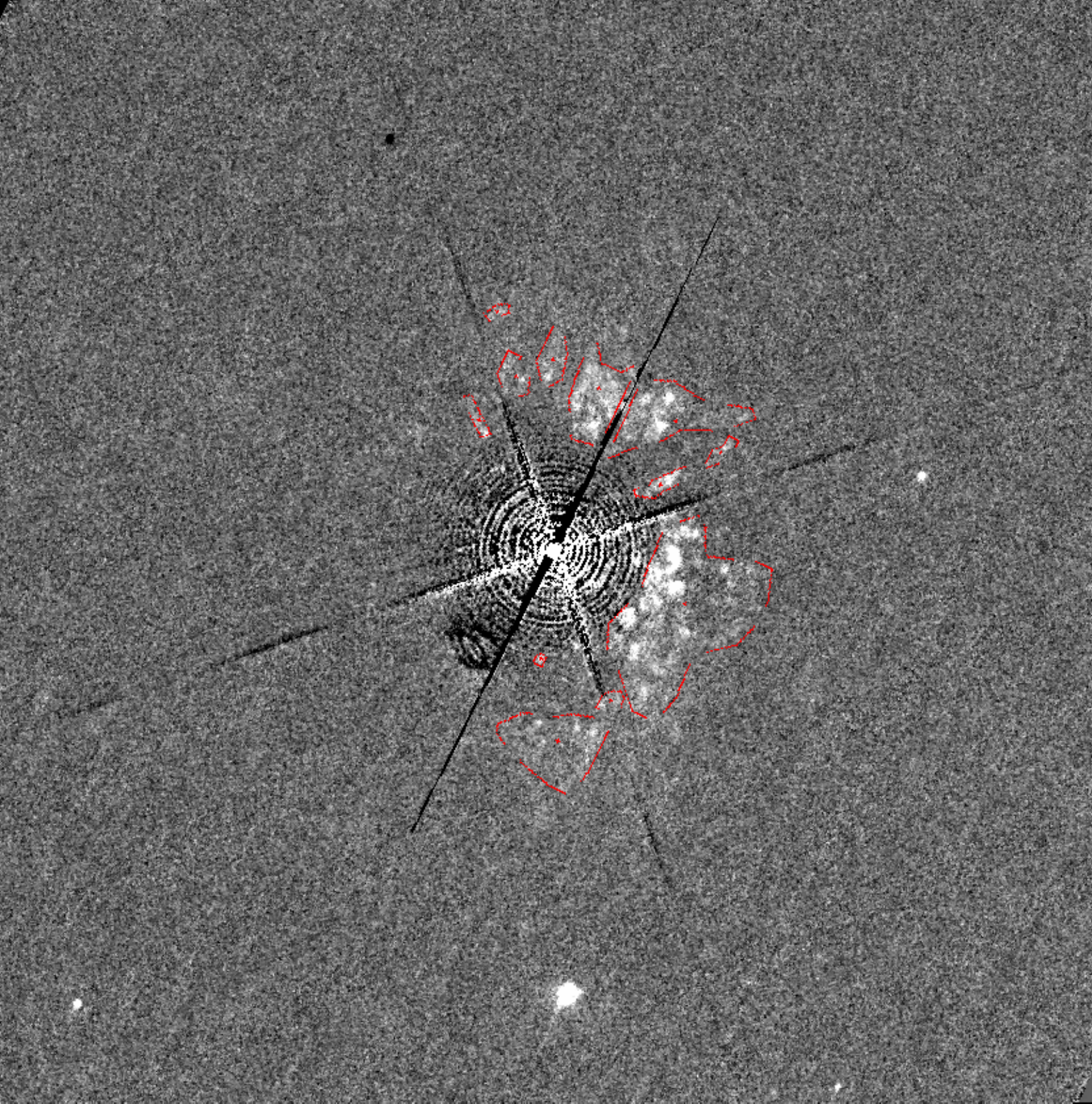}

\caption{Closeup view of the PSF-subtracted F656N images of T Pyx taken by {\it HST} on 2011 November 15. North is up, East is left and the image is $30"$on each side. The encircled areas highlight the H$\alpha$ emission. Almost all of the H$\alpha$ emission seen on 2011 September 25, north and south of T Pyx, is gone, and replaced by emission northwest and southwest of the star.}

\end{figure}

\clearpage

\begin{figure}

\figurenum{7}

\epsscale{1.0}

\plotone{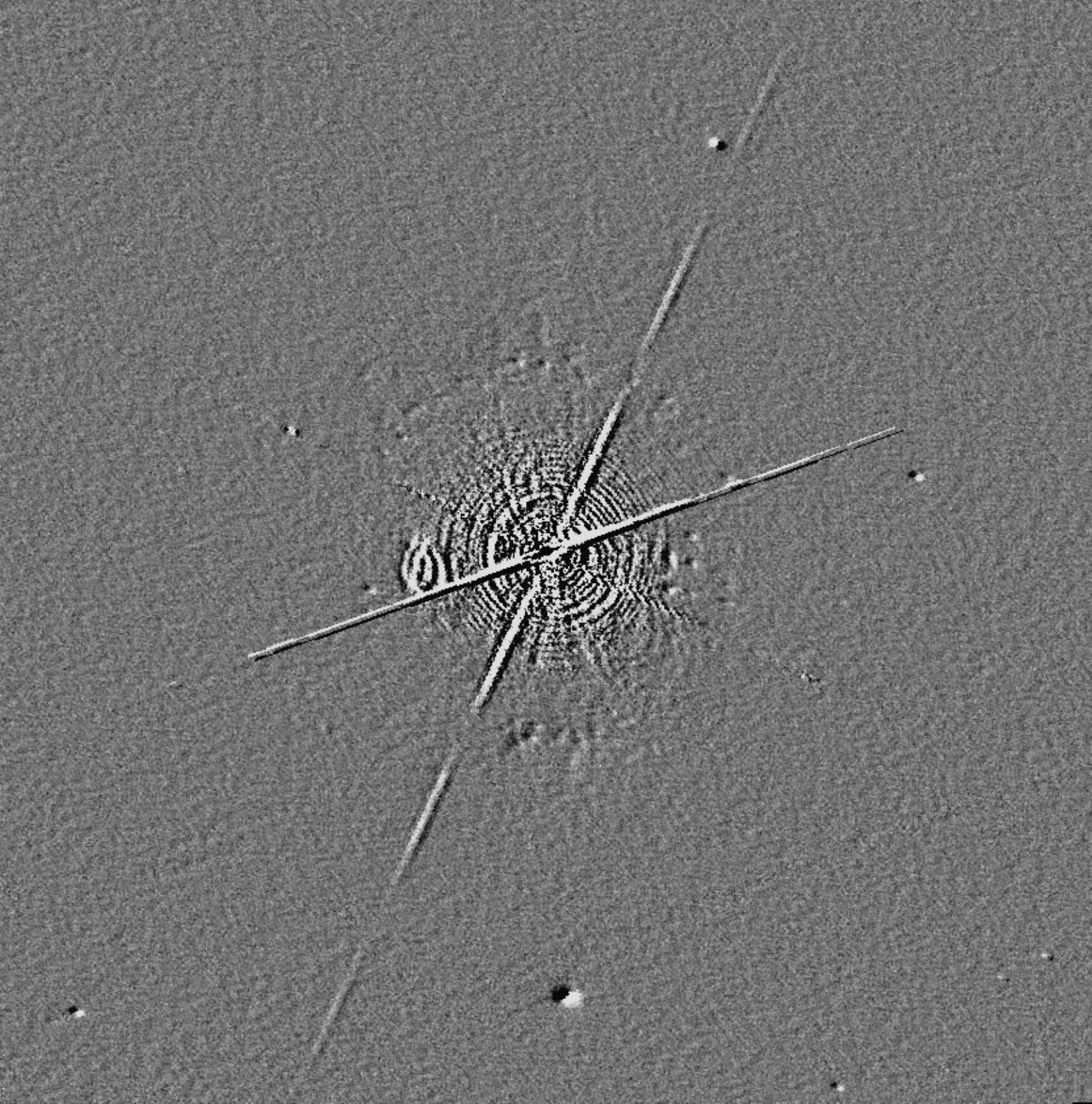}

\caption{An unsharp mask of the PSF-subtracted F656N image of T Pyx taken on 2011 September 25. North is up, East is left and the image is $30"$ on each side. The presence of just a few black-white paired point-like features emphasizes that the H$\alpha$ emission on this date is mostly diffuse rather than point-like.}

\end{figure}

\clearpage

\clearpage

\begin{figure}

\figurenum{8}

\epsscale{1.0}

\plotone{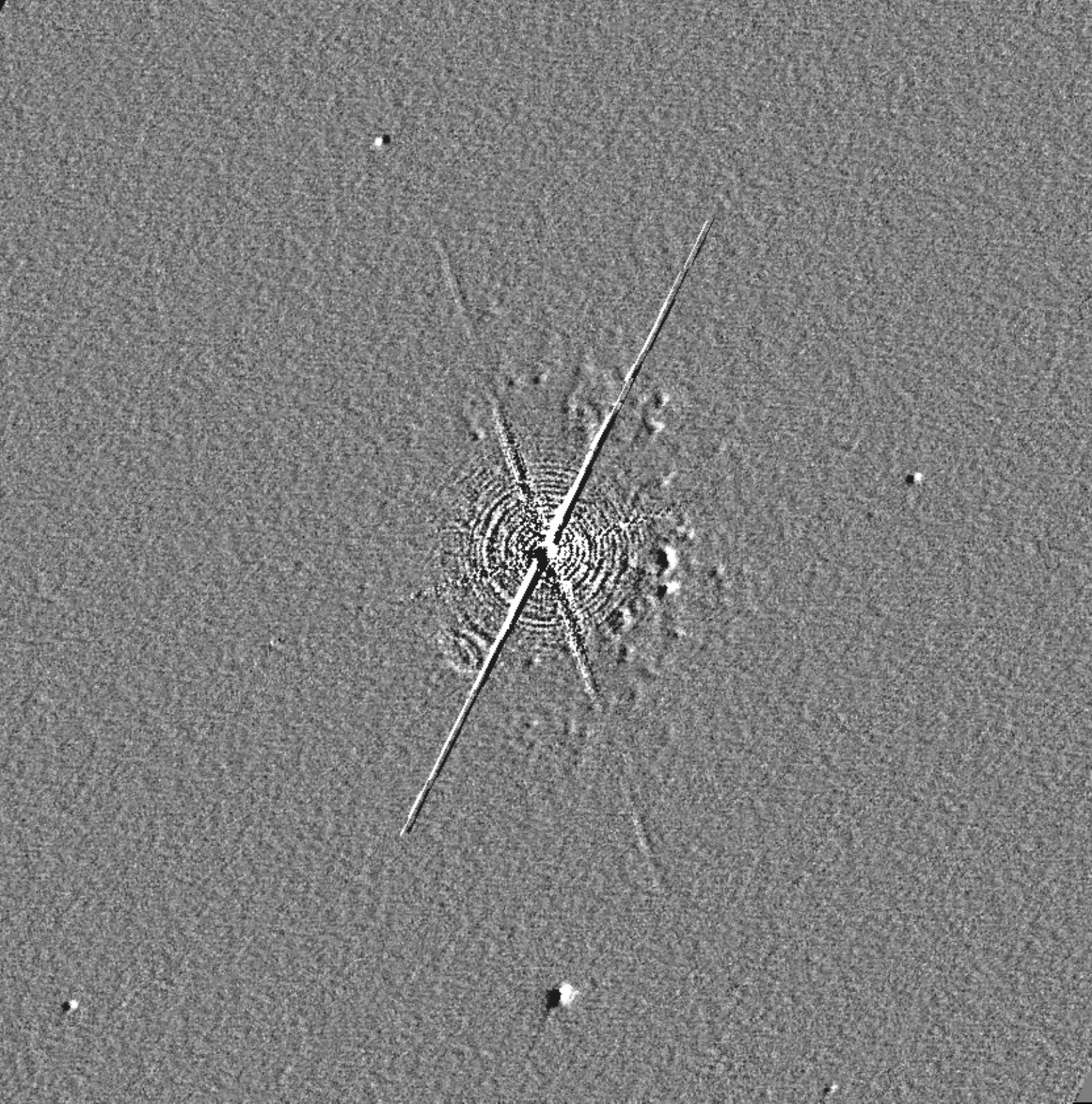}

\caption{An unsharp mask of the PSF-subtracted F656N image of T Pyx taken on 2011 November 15. North is up, East is left and the image is $30"$ on each side. This again emphasizes that the H$\alpha$ emission on this date is mostly diffuse rather than point-like.}

\end{figure}

\clearpage

\clearpage

\clearpage

\begin{figure}

\figurenum{9}

\epsscale{1.0}

\plotone{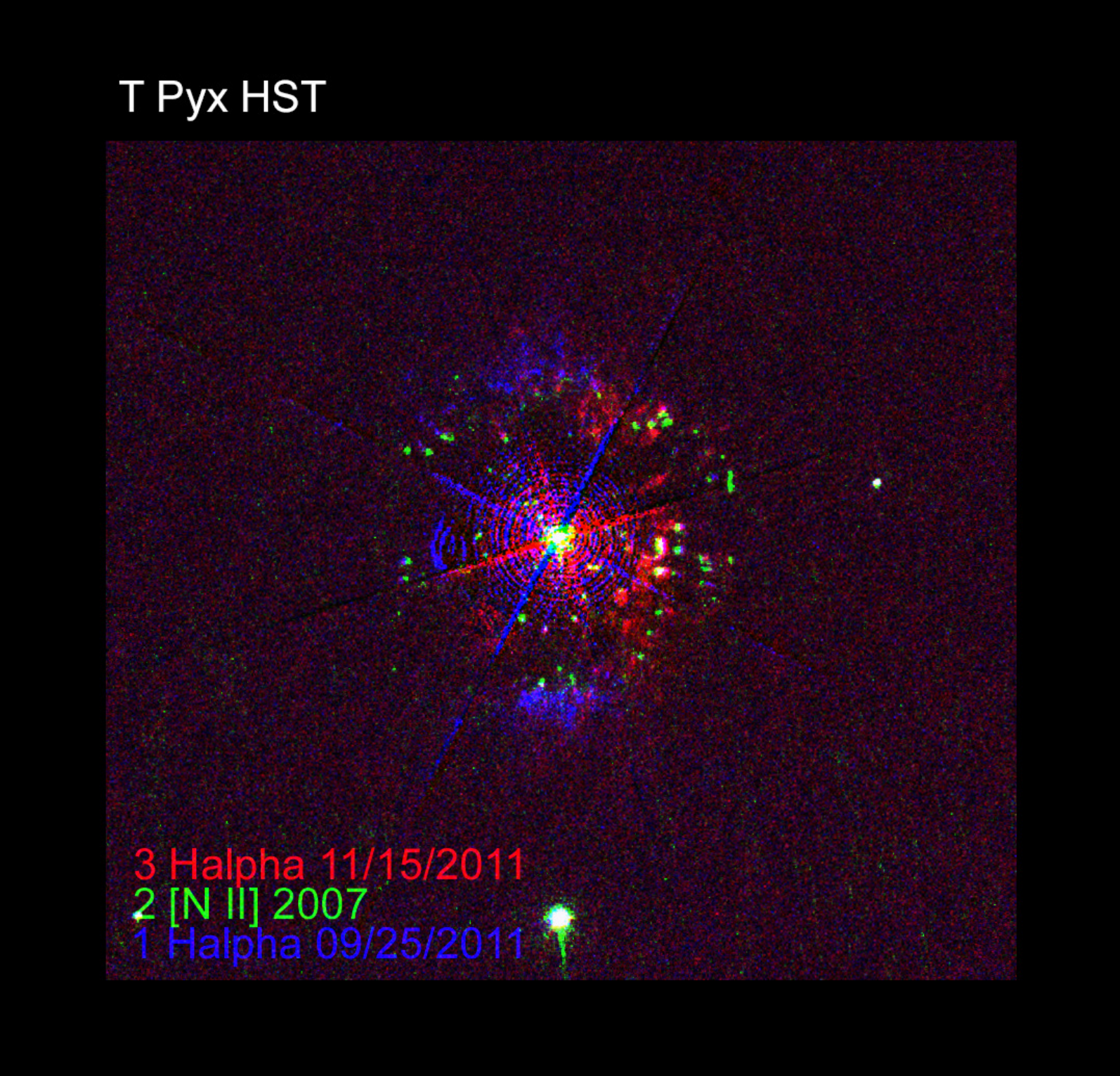}

\caption{The sum of the two epochs of strongest H$\alpha$ emission ({\it HST} F656N  images of T Pyx taken on 2011 September 25 (in blue) and 2011 November 15 (red) ), superposed on the most recent archival {\it HST} [\ion{N}{2}] F658N image (in green) taken in 2007. North is up, East is left and the image is $30"$ on each side. The [\ion{N}{2}] emission is concentrated in unresolved knots, while the H$\alpha$ emission is largely diffuse.}

\end{figure}

\clearpage

\section{Results}

\subsection{More Distant Ejecta}

The key science question motivating our {\it HST} observations of T Pyx is the presence or absence of previously undetected ejecta. Figures 4, 5, and 6 make it clear that we have discovered substantial amounts of previously invisible, cold hydrogen ejecta, briefly illuminated and photoionized by the eruption of T Pyx. Features in the PSF-subtracted images on 2011 April 30 and 2011 July 08 to the north, east and southeast are artifacts - internal reflections - produced by the H$\alpha$ filter. However, on 2011 September 25, there was a dramatic brightening in H$\alpha$ emission to the north and south of T Pyx. Figure 7 demonstrates that the H$\alpha$ emission is much more diffuse and less clumpy than the [\ion{N}{2}] emission seen in the same region in 2007. Just as significant, there is no trace of H$\alpha$ emission outside of the region of [\ion{N}{2}] emission \citep{sha97,sps10}. 

Fifty one days later, on 2011 November 15, the photoionized H$\alpha$ emission to the north and south of T Pyx was gone. In its place there was extensive H$\alpha$ emission to the northwest, west and southwest of T Pyx. This H$\alpha$ emission overlaps regions of [\ion{N}{2}] emission seen in 1997 and 2007. There was, again, no trace of H$\alpha$ emission outside the regions of [\ion{N}{2}] seen in 1997 and 2007. The locations of these transient H$\alpha$ detections overlap the dusty patches detected (with low S/N) in the continuum images of \citet{sok13}. They are consistent with the conclusions of \citet{sok13} that T Pyx's ejecta are concentrated in a ring about $5"$ in radius, tilted 30 - 40 degrees with respect to the plane of the sky, with the eastern edge tilted toward the observer, at a distance of $ 4.8 \pm 0.5$ kpc. 

Even more remarkable are the H$\alpha$ images of 2012 February 22 and 2012 May 01. Almost all of the H$\alpha$ emission that was so prominently visible on 2011 November 15 has been extinguished. The only exceptions are two unresolved knots located $4"$ west of T Pyx on 22 Feb 2012, which are still seen on 01 May 2012. {\it No new H$\alpha$  emission appeared outside the regions of [\ion{N}{2}] emission}, even though sufficient time has elapsed for the burst of nova light to photoionize hydrogen twice as far away from T Pyx as was detected in 2011 September and November, if it were present in the ring-like geometry suggested by \citet{sok13}. 

To put limits on the surface brightness of H$\alpha$ emission outside the [\ion{N}{2}]-dominated ejecta, we have constructed, in Figure 10, a montage of four images (labeled A, B, C and D), with successively fainter artificial nebulosities. The region outlined with a red rectangle in the figure labeled A encloses the nebula which, appropriately scaled, we transplanted to the region indicated by the blue ellipse. The centroid of the blue ellipse is (arbitrarily) twice as far as the centroid of the red rectangle from T Pyx. The precise position of the blue ellipse is unimportant, as the background surface brightness is constant outside the limits of the T Pyx nebulosity. In image A, the flux in the red rectangle is simply added to the pixels in an identically-sized rectangular region inside the blue ellipse. This addition has the effect of increasing the noise, which is why the rectangular shape of the synthetic nebulosity can be discerned. In figure B we reduced the flux from the red rectangle region by a factor of two and again added the synthetic nebulosity to the pixels in the blue ellipse. In figures C and D we repeated this procedure, after reducing the flux by factors of 4 and 8, respectively. Reductions by factors of 2 and 4 (images C and D) made the synthetic nebulosity fainter, but it remained detectable. In image D it is clear that the background noise dominates the signal, and the synthetic nebulosity with 1/8 the surface brightness of the original is just at the limits of detectability. 

\clearpage

\begin{figure}

\figurenum{10}

\epsscale{0.85}

\plotone{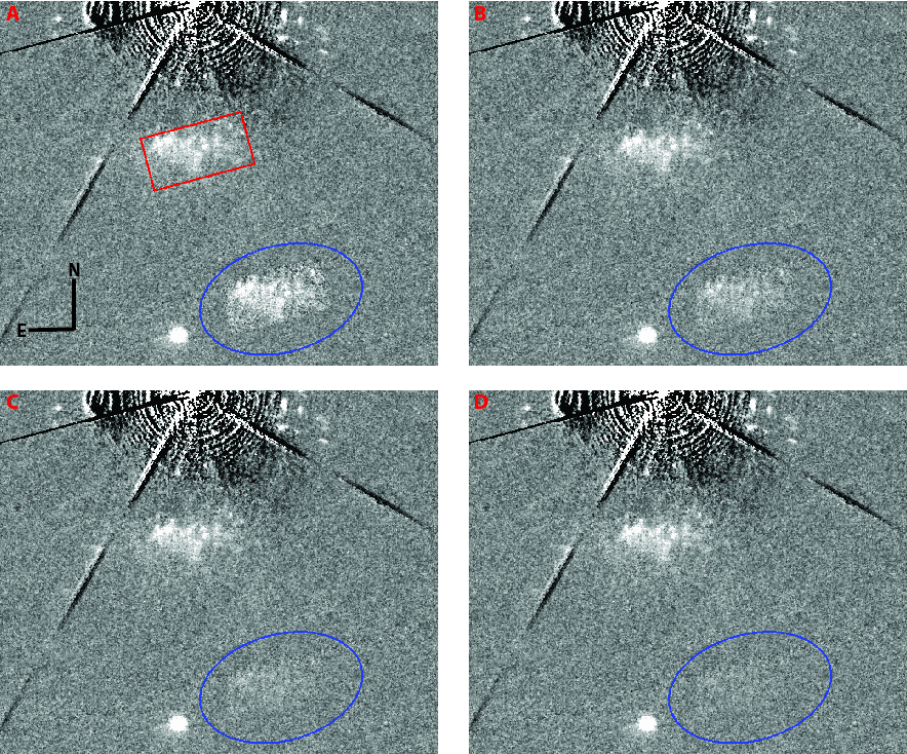}

\caption{A montage of four images (labeled A, B, C and D) of the region next to Pyx. We have copied the flux, pixel-by-pixel, of the nebulosity inside the red rectangle of image A. Synthetic nebulosities with 100\%, 50\%, 25\% and 12.5\% of the detected flux in the red rectangle were added inside the blue ellipse in images A, B, C and D, respectively. See the text for more details.}

\end{figure}

\clearpage

We conclude that there {\it is} a large reservoir of previously undetected, cold hydrogen mixed in with the extensive filamentary [\ion{N}{2}] structures first identified in \citet{sha97}. However, no extended reservoir of cold hydrogen, comparable in mass to that which we have now discovered, exists out to roughly twice the distance from T Pyx (near the plane of the sky) in which [\ion{N}{2}] knots and the newly detected H$\alpha$ emission are seen. The lack of such extended hydrogen is strong evidence against the hypothesis that T Pyx has been ejecting matter as a RN for many centuries or millennia.

\subsection{Mass of Previously Unseen Ejecta}

We were previously able to put a hard lower limit on the ejecta mass of T Pyx of 1.6 x $10^{-6}M_\odot$ via the [\ion{N}{2}] flux, the upper limit on the knots' electron density to allow the forbidden [\ion{N}{2}] lines, and the number and size of emitting knots \citep{sha97}. Correcting for the larger distance (4.8 kpc) that is now deduced for T Pyx (Sec. 1), the lower limit on the mass in [\ion{N}{2}]-emitting knots is 1.6 x $10^{-5}M_\odot$, consistent with the mass expected from a classical nova eruption. \citet{sps10} demonstrated via simulations that this is still an underestimate, by a factor of roughly 4.4, because knots are turning on and off as successive generations of ejecta collide with each other and then recombine. The ejecta mass seen in [\ion{N}{2}] is thus about 7 x $10^{-5}M_\odot$.

The rapid disappearance of the H$\alpha$ emission to the north and south of T Pyx that we see in the 51 days between 2011 September 25 and 2011 November 15 allows us, for the first time, to put a lower limit on the density of hydrogen photoionized by the T Pyx eruption. The e-folding time $t_{\rm rec}$ over which electrons recombine in ionized hydrogen is $10^{5}$/$N_{\rm e}$ yr, where $N_{\rm e}$ is the number density of free electrons \citep{ost89}. \citet{sps10} note that over a time $t_{\rm rec}$ in recombining hydrogen, $N_{\rm e}$ falls by a factor of 2.7, the total flux from a region of gas falls by a factor of 7.4, corresponding to 2.2 mag. This is what we measure for the fading of the hydrogen to the north and south of T Pyx to within $\pm$ 20\%.  Thus that hydrogen must possess a density of at least 7 x $10^{5}$ $\rm  cm ^{-3}$ . This lower limit is an order of magnitude larger than the largest density of nitrogen that is consistent with the strong [\ion{N}{2}] lines in the knots of T Pyx. The similar, total disappearance of emitting hydrogen northwest and southwest of T Pyx between 2011 November 15 and 2012 February 22 (a 101 day interval, from 282 to 383 days after eruption) allows us to place a lower limit of $ 3.5 \times 10^{5}$  $\rm cm^{-3}$ on that material.

A Space Telescope Imaging Spectrograph (STIS) spectrum of the emitting material between $4"$ and $6"$ northwest of T Pyx was obtained by \citet{sok13} on 2011 December 5 (see their Figure 2). Two emission lines are clearly evident: H$\alpha$ and [\ion{O}{3}] $\lambda$5007, with comparable equivalent widths. The presence of strong 
[\ion{O}{3}] $\lambda$5007 emission puts an upper limit on the density of 7 x $10^{5}$ $\rm cm ^{-3}$ to avoid collisional de-excitation \citep{ost89}. This is similar to the lower limit obtained in the previous paragraph.

It is tempting to suggest that this density, higher by almost an order of magnitude than that which can exist in the [\ion{N}{2}] knots, in a surface area similar to that occupied by the knots, represents roughly 10 times more mass than has previously been deduced to exist in T Pyx's ejecta, viz. 7 x $10^{-4}M_\odot$ in total. The dynamical, radio and X-ray estimates of the mass ejected in the latest outburst ($3 \times 10^{-5}M_\odot$ or larger) are certainly consistent with our total mass estimate of 7 x $10^{-4}M_\odot$. Consistency is maintained if all six observed eruptions (and the deduced 1866 eruption) ejected $3 \times 10^{-5}M_\odot$ . This quantity of ejected mass is usually associated with a classical, rather than a recurrent nova. It is no longer clear that the last six eruptions are different from the first eruption of 1866.

\citet{tor13} predicted that the bright knots seen in [\ion{N}{2}] would be connected by a filamentary network of fainter material as a consequence of the growth of Rayleigh-Taylor instabilities in the colliding ejecta shells. The flash ionized material reported in this paper has morphology and spatial distribution consistent with their prediction. However, the input parameters for the ejected mass in the simulations by \citet{tor13} were based what was then known about T Pyx's distance and the [\ion{N}{2}] images.  That yielded shell densities two orders of magnitude below the densities derived here, suggesting that further modeling will be needed to make a useful quantitative comparison. Nevertheless, the agreement in morphology lends support to the above suggestion that the ejecta are higher in mass, by an order of magnitude, than previously deduced.

\section{Summary and Conclusions}

Seven epochs of {\it HST} WFC3 F656N narrowband H$\alpha$ images of the RN T Pyx were obtained during the year following its 2011 eruption. Significant ionization and fluorescence of hydrogen ejecta was seen 132 days and 183 days after outburst, but not earlier or later. The rapid recombination of the photoionized hydrogen 
and presence of [\ion{O}{3}] $\lambda$5007 demonstrates densities, and thus probably ejecta masses, an order of magnitude larger than previous estimates. The morphology of the flash-ionized hydrogen is similar to that predicted by the hydrodynamic models of \cite{tor13}, viz. a filamentary network of fainter material connecting the bright [\ion{N}{2}] knots, produced by the growth of Rayleigh-Taylor instabilities in colliding ejecta shells. The lack of detections of hydrogen later than 183 days after eruption rules out ejecta of similar density and mass to about twice the distance of any currently known material. The absence of more distant ejecta supports the scenario wherein T Pyx has been erupting only in the past 150 years, after many millennia of quiescence.

\acknowledgments

We gratefully acknowledge the support of the STScI team responsible for ensuring timely and accurate 
implementation of our T Pyx ToO program. We especially thank Lisa Frattare for creating the multi-epoch image that is Figure 8. Support for program \#12446 was provided by NASA through a grant from the Space Telescope Science Institute, which is operated by the Association of Universities for Research in Astronomy, Inc., under NASA contract NAS 5-26555. We also thank M. Linnolt and E. Waagan of the AAVSO for rapidly bringing T Pyx's latest eruption to the attention of the astronomical community.

\end{document}